\documentclass[prb,twocolumn,showpacs,showkeys,amsmath,amssymb]{revtex4} %Physical Review B twocolumn.

\usepackage{graphicx}% Include figure files
\usepackage{dcolumn}% Align table columns on decimal point
\usepackage{bm}% bold math
\usepackage{bbm}
\newcommand{\mb}{\mathbf}

\begin{document}

\preprint{APS/123-QED}

\title{A Monte Carlo study of surface sputtering by dual and rotated ion beams}

\author{Taha Yasseri}
\email{yasseri@theorie.physik.uni-goettingen.de}
\homepage{http://www.theorie.physik.uni-goettingen.de/~yasseri}
\affiliation{Institut f\"{u}r Theoretische Phyisk, Friedrich-Hund Platz 1, D-37077 G\"{o}ttingen, Germany.}

\author{Reiner Kree}
\email{kree@theorie.physik.uni-goettingen.de}
\affiliation{Institut f\"{u}r Theoretische Phyisk, Friedrich-Hund Platz 1, D-37077 G\"{o}ttingen, Germany.}

\date{\today}

\begin{abstract}
Several, recently proposed methods of surface manufacturing based on
ion beam sputtering, which involve  dual beam setups, sequential
application of ion beams from different directions, or sample rotation, are studied with the method
of kinetic Monte Carlo simulation of ion beam erosion and surface
diffusion. In this work, we only consider erosion dominated
situations. The results are discussed by comparing them to a number of
theoretical propositions and to experimental findings.   
Two ion-beams aligned opposite to each other produce stationary,
symmetric ripples. Two ion beams crossing at right angle will produce square patterns
only, if they are exactly balanced. In all other cases of crossed
beams, ripple patterns are created, and their orientations are shown
to be predictable from linear continuum theory. In sequential ion beam
sputtering we find a very rapid destruction of structures created from the
previous beam direction after a rotation step, which leads to a transient
decrease of overall roughness. Superpositions of patterns from several
rotation steps are difficult to obtain, as they exist only in very
short time windows. In setups with a single beam directed towards a
rotating sample, we find a non-monotonic dependence of roughness on
rotation frequency, with a very pronounced minimum appearing at the
frequency scale set by the relaxation of prestructures observed in
sequential ion beam setups. Furthermore we find that the logarithm of the height
of structures decreases proportional to the inverse frequency.
\end{abstract}


\maketitle

%%%%%%%%%%%%%%%%%%%%%%%%%%%%%%%%%%%%%%%%%%%%%%%%%%%%%%%%%%%%%%%%%%%%%%%%%%%%%%%%%%%%%

\section{\label{sec:Intro}Introduction}
In recent years ion-beam sputtering (IBS) as a method of nanoscale surface manufacturing \cite{navaez62} has attracted 
much experimental work (for a recent review see Ref.~\onlinecite{frost08}). 
Self-organized ripple patterns appear generically by oblique incidence single-beam IBS, and dot or hole patterns are obtainable by
normal-incidence IBS or by oblique incidence IBS on rotating samples.  To produce a larger variety of structures and to improve their
quality,  more complex setups with multiple ion beams, IBS on prestructured samples, 
and rotating samples have been used, but few of them have been investigated in detail. 
In particular, Carter \cite{carter04,carter05,carter06} has proposed the use of dual ion beam sputtering (DIBS), Vogel and Linz 
\cite{vogel07} proposed a four-beam setup and
claimed that corresponding results may be obtained from a sequence of prestructuring and stepwise beam or sample 
rotation using a single ion beam (sequential ion beam sputtering or SIBS). 
Continuous rotation of the sample or the ion beam (referred to as RIBS, i.e. rotating IBS, in the following) has been proposed to 
suppress ripple formation \cite{zalar85,zalar86} (for example in SIMS and AES, 
where ripple formation would reduce the depth profiling resolution), but also to enhance the quality of isotropic structures.\cite{bradley96a} 
These proposals were based on continuum theories of the self-organized pattern formation by IBS, which are further developments 
of the work of Sigmund \cite{sigmund69} , Bradley and Harper  \cite{bradley88} and  Makeev et al..\cite{makeev02} Recently, Joe, Kim et al.\cite{joe09} conducted a 
systematic experimental study of DIBS and SIBS on Au(001).  A number of their findings are not in 
accordance with expectations derived from the conventional continuum framework.  This motivated us to study DIBS, SIBS and RIBS setups
with Monte Carlo (MC) simulation methods. Recent proposals \cite{vogel07,munoz-garcia09} are based upon extensions of the standard 
continuum model (which is formulated as an anisotropic and noisy  Kuramoto-Sivashinsky equation). These extensions have introduced new physical 
mechanisms, which change the scenario of pattern formation of the standard model, but none of these mechanisms has been confirmed 
and tested independently beyond doubt. In this situation, it may be helpful to see, what can be achieved from a simulation, which is a straightforward
stochastic implementation of Sigmund's energy deposition and sputtering formula, combined with independently tested models of
surface diffusion in simple model systems, but without any further approximations.

In the next section, we briefly introduce our MC simulation method  and the geometries of the considered setups. 
Subsequently, we present and discuss results we obtained for the topographies, the shape and orientation of the ripples, 
the structure function and the evolution 
of roughness in different setups. We compare our findings to theoretical proposals, experimental results and to the standard linear continuum model.

\section{\label{sec:MC}Monte Carlo simulation}
Throughout this work, we will
focus on the erosion dominated regime, leaving more complex interplays between erosion and 
diffusion in multi-beam and rotating setups for further studies. 
Parameters of the simulations (given below) are chosen appropriately, such that 
in single ion beam setups,  ripples
perpendicular to the direction of the beam are produced on non-rotating samples  and cellular structures occur on rotating samples.\cite{yewande06, yewande07}

\begin{figure}
(a)\includegraphics[width=.43\linewidth]{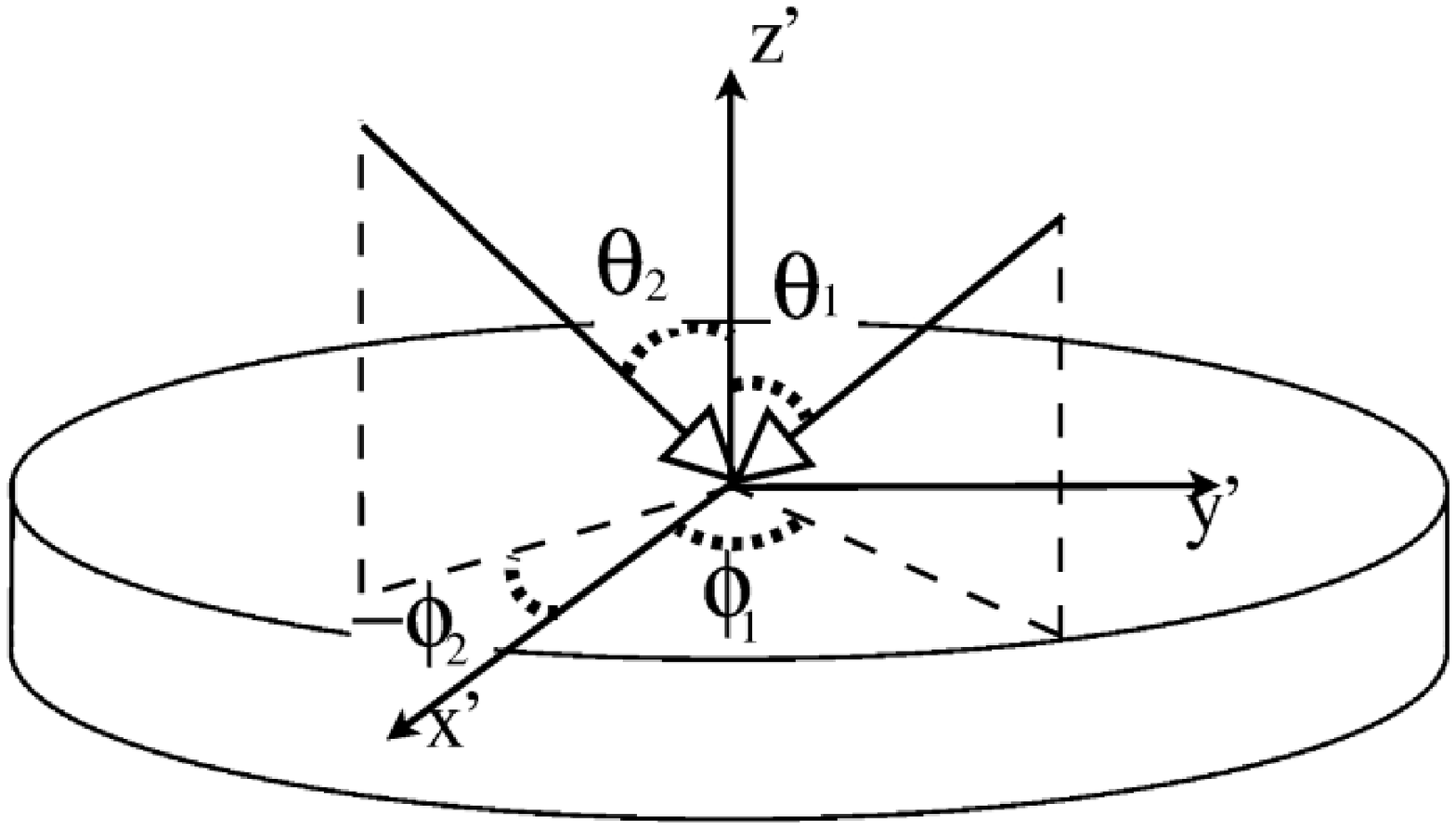} (b)\includegraphics[width=.43\linewidth]{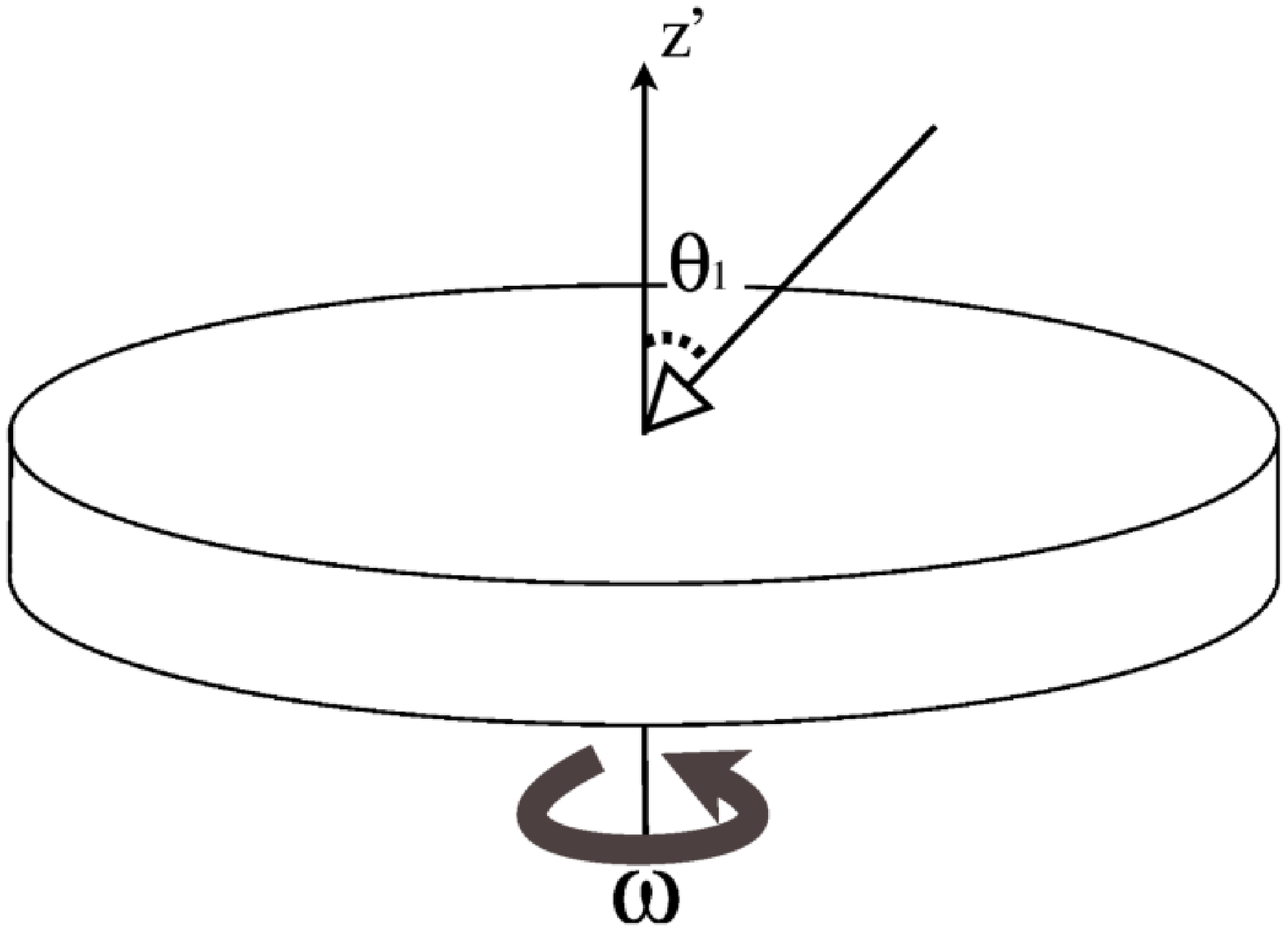}
(c)\includegraphics[width=.95\linewidth]{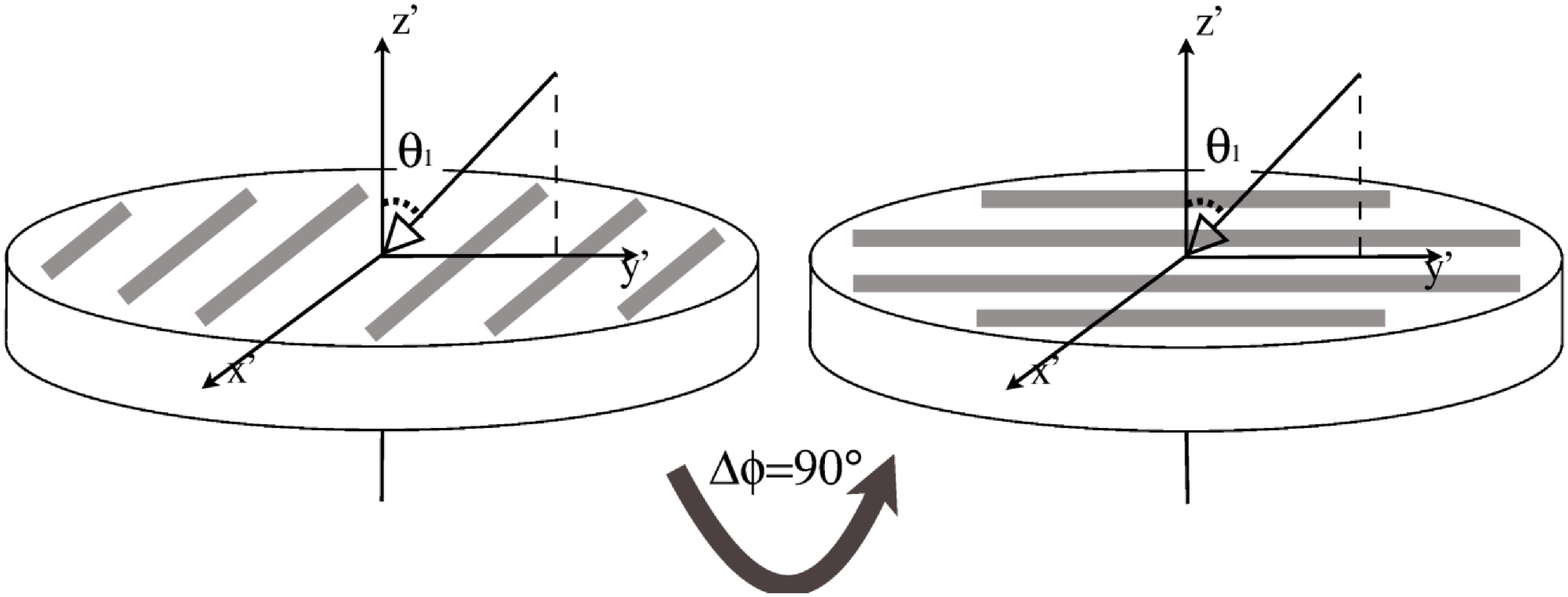}
\caption{\label{fig:set-up} Three different IBS experiment setups. (a)
  dual ion beam with fixed sample (DIBS).
 (b) continuously rotating sample with single ion beam (RIBS). 
(c) sequentially sputtering of pre-rippled surfaces after $90^\circ$
rotation (SIBS).}
\end{figure}

Fig.~\ref{fig:set-up} depicts the geometries of setups we use in the simulations.  A DIBS setup consists of two ion beams incident from directions 
described by polar angles $\theta_1$ and $\theta_2$ and azimuthal 
angles $\phi_1$ and $\phi_2$ (see Fig.~\ref{fig:set-up}(a)). 
As a simple special case, we will consider opposing beams, i.e. $\theta_1=\theta_2$ and $\Delta \phi=\phi_1-\phi_2=180^\circ$. Crossed beams  
are studied for equal polar angles as well as for the general case
 of different polar  and azimuthal angles.

SIBS setups are shown in  Fig.~\ref{fig:set-up}(c).  In a first step ripples are produced by a single ion
beam. Then  we change the azimuthal angle of the ion beam direction by some $\Delta\phi$  and monitor the further evolution  
of surface structures.

Finally, Fig.~\ref{fig:set-up}(b) shows the RIBS setup, which is characterized by a constant angular velocity
$\omega$ of the sample, which we realize by a fixed sample and a correspondingly rotation beam. The evolution of structures and surface roughness will be systematically studied as a function of $\omega$. 

A first version of our Monte Carlo model of erosion and surface diffusion has been introduced in Ref.~\onlinecite{hartmann02}. Further developments of the model are discussed in Ref.~\onlinecite{hartmann09}.
Here, we use a solid-on-solid cubic lattice with an initially flat surface of size  $L\times L$ with $L=512$. 
Its configuration is described by a single valued, time dependent, height function $h(x,y,t)$, which obeys periodic boundary
conditions.

Ions start from a random position above the surface and move towards the surface in direction  $(\theta,\, \phi)$. 
Erosion occurs as a result of energy transfer via collision cascades induced by impinging ions. 
The ions penetrate the surface through a depth of $a$.
The average energy distribution of a cascade 
initialized by a single ion is approximated by a Gaussian distribution which has width of $\sigma$ 
in directions of ion-beam and $\mu$  perpendicular to the ion-beam, respectively. 
As default values, we use $a=9$, $\sigma=3$ and $\mu=1.5$ lattice constants.
Every lateral atom is a candidate for erosion, the probability of erosion being proportional 
to the amount of energy, which reaches the atom. The parameters are tuned so that the average yield is $\sim 7$ atoms per ion. The natural time scale in this type of MC simulations is proportional to ion fluence, therefore we use
1 ML $=$ $L^2$ ions, corresponding to 1 ion per atom at the flat surface as unit of time in this work. It is connected to laboratory time via the ion flux.

Surface diffusion is modeled as thermally activated hopping. In a diffusion sweep every surface atom  
has the chance to hop to a nearest neighboring site. The attempt frequency is calibrated to a temperature of $T\simeq 400 K$ 
and a substrate barrier of $E_s=0.75 eV$ is chosen. This implies that after every $L^2/1000$ incoming ions, 
one diffusion sweep over the surface is taken.\cite{hartmann09}
The hops are accepted with the probability of $\exp(-\Delta E /k_B T)$. The additional energy barrier ($\Delta E$) consists of two terms,
$\Delta E = nE_{nn}+E_{SB}$. Here $n$ is the net number of broken bonds after the hop. 
If this net number is negative, i.e. additional bonds have been created, $n$ is put equal to zero. We choose $E_{nn}=0.18 eV$ 
as nearest neighbor bond energy. $E_{SB}=0.15 eV$ is an Ehrlich-Schwoebel barrier term (ES), which is only nonzero, if an ad-atom 
approaches a step edge from an upper terrace. More details about the surface diffusion modeling can be found in Ref.~\onlinecite{hartmann09}.

%\subsection{\label{sec:Rotation}Sample rotation}
The rotation of sample during IBS is equivalent to a fixed sample and a rotating ion beam. 
Therefore rotation or any change of azimuthal angles in lab coordinates is simulated by keeping the 
surface fixed and rotating the ion beam correspondingly.

In a single beam setup, a typical run of the above described MC simulation, starts from a flat surface, after $t\sim3 $ the first, short ripples appear, and after 
$t\sim 6 $ regular ripples of the size of the system have formed (see Fig.~\ref{fig:normal-opposed}(a)). 
As fluence increases, the ripple pattern shows less defects (branching and deviation from expected direction). 
We also observe ripple motion, but will not report details about its characteristics in this work.\cite{yewande05}

\section{\label{sec:Results}Results and Discussion}

\subsection{\label{sec:Opposed}Opposed ion-beams}
It has been proposed  by Carter \cite{carter05} that IBS with two diametrically opposed ion-beams 
(same $\theta$ and a difference of $180^\circ$ in $\phi$) can lead to a cancellation of instabilities induced by each  beam. 
Instead, he predicted temporal oscillations of ripple amplitudes. 
Furthermore,  ripple motion should be suppressed due to the restoration of reflection symmetry, which would be broken by a single beam.

\begin{figure}
(a)\includegraphics[width=.43\linewidth]{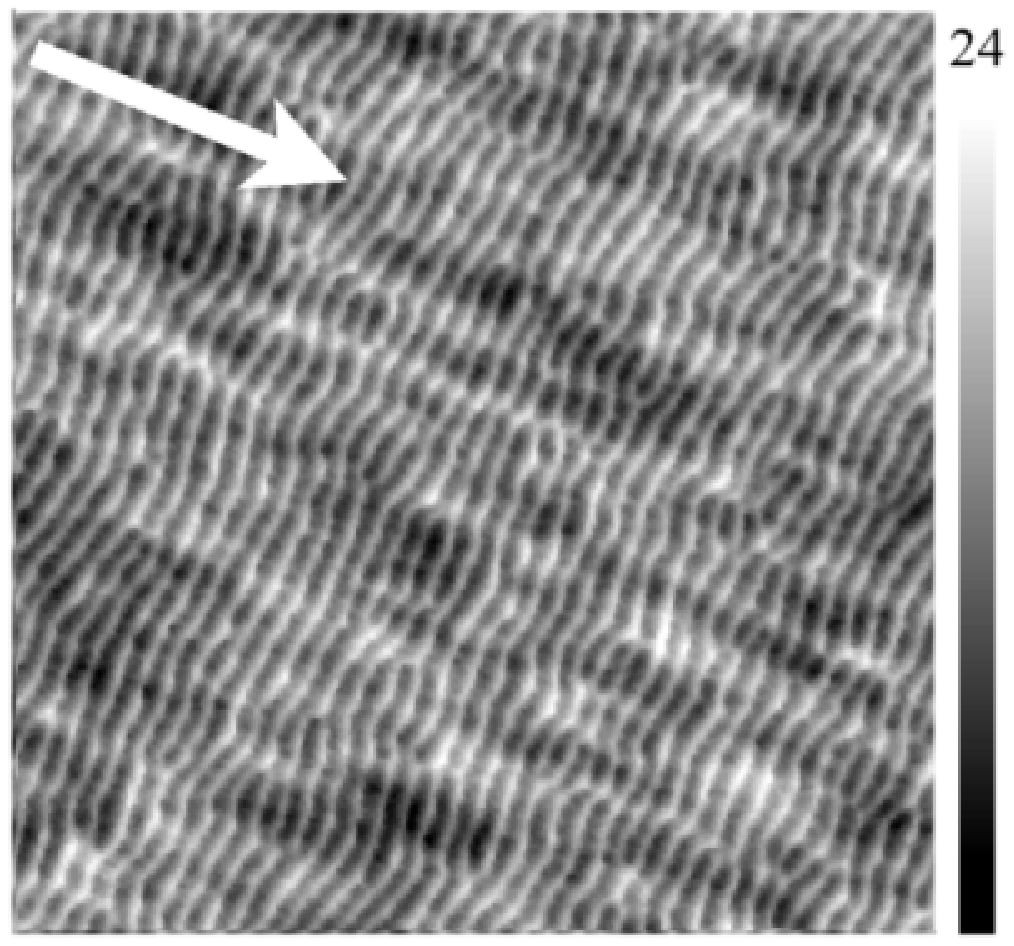} (b)\includegraphics[width=.43\linewidth]{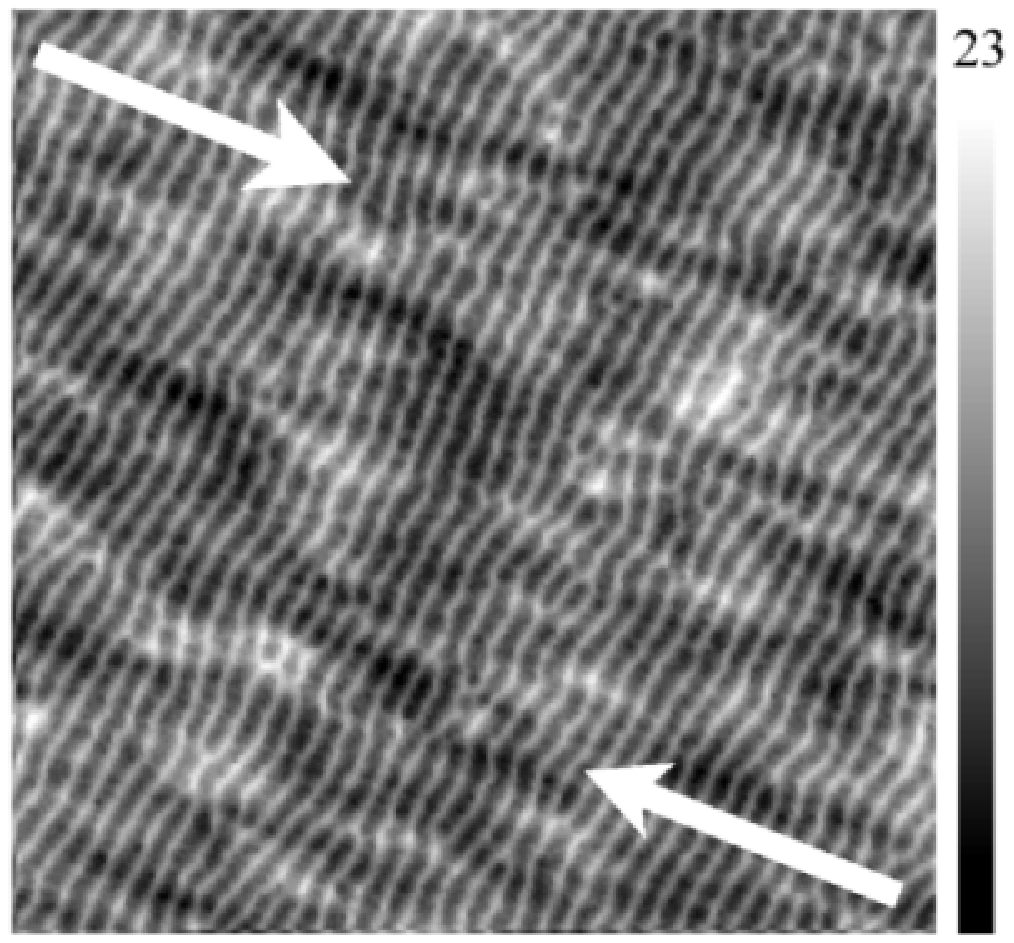}
\caption{\label{fig:normal-opposed} Rippled surfaces after $8 ML$ of
  sputtering. (a) by a single ion beam, (b) by two ion beams opposed
  to each other. Arrows indicate the direction of  ion beams.}
\end{figure}

\begin{figure}
\includegraphics[width=\linewidth]{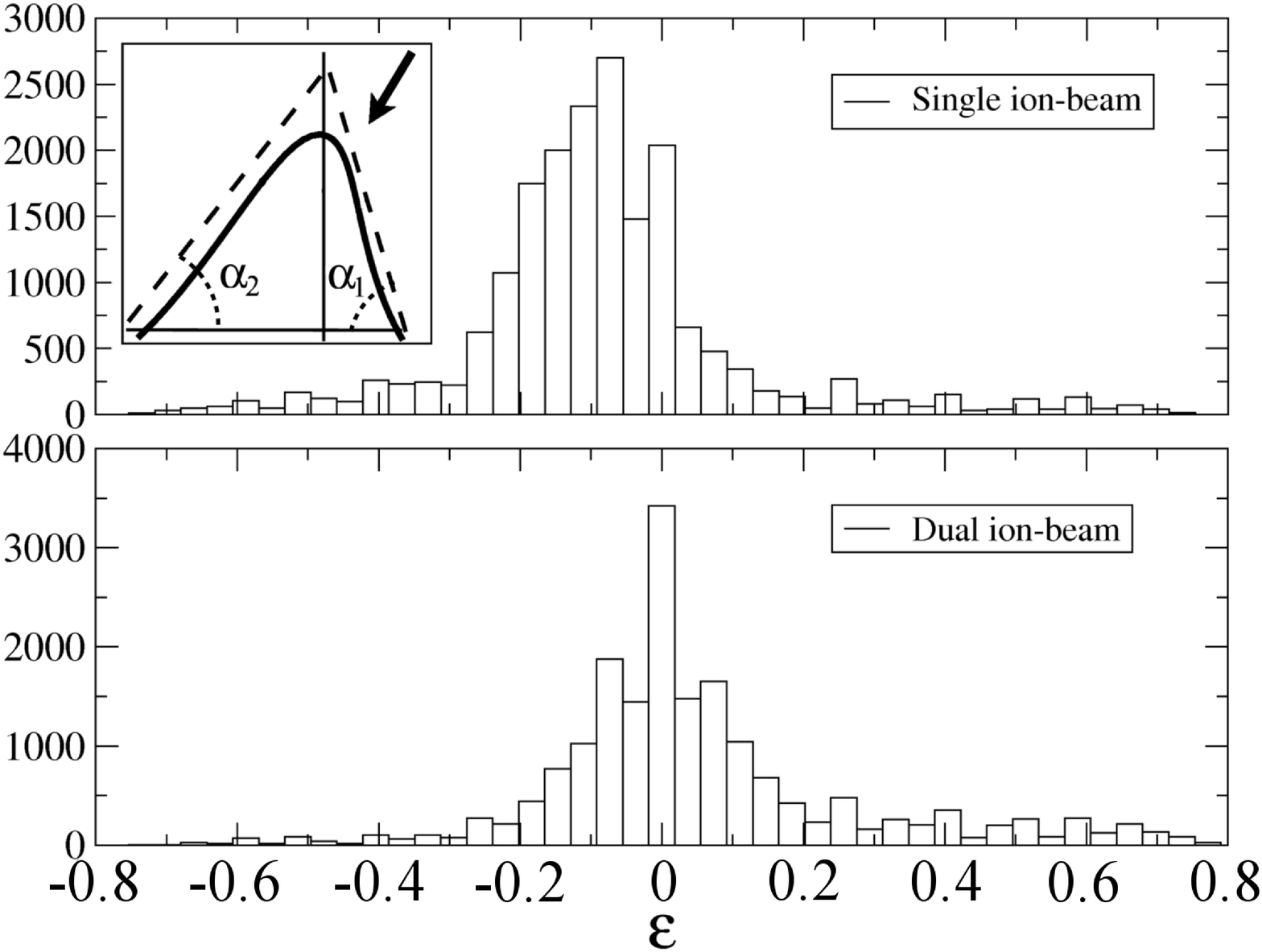}
\caption{\label{fig:epsilon} Distribution function of the asymmetry parameter $\epsilon$ (see main text) for the cases of a single ion
  beam (upper panel) and two ion beams opposed to each other (lower panel). }
\end{figure}

 Fig.~\ref{fig:normal-opposed}(b) shows results of topographies obtained from MC simulations of this setup. 
 We never observe the predicted 
 behavior, instead the growth of structures resembles that obtained in a single beam setup, as can be seen from comparing Figs.~\ref{fig:normal-opposed} (a) and (b). 
But ripples obtained from opposed beams appear longer and more straight than those from single beams and the patterns contain less defects. Furthermore, the
shape of ripples changes significantly towards more symmetric slopes.   
In Fig.~\ref{fig:epsilon}, we show a quantitative analysis of ripple slope angles $\alpha_1$ and $\alpha_2$, which are defined in the inset of the upper panel.
To measure the symmetry of the slopes, we consider the asymmetry parameter
\begin{equation}
\epsilon=\frac{\alpha_2-\alpha_1}{\alpha_2+\alpha_1}
\end{equation} 
Histograms of this quantity are shown in Fig.~\ref{fig:epsilon} for single beam and opposed beam setups. The average value
of $\epsilon$  is $\bar{\epsilon}\approx -0.08$ for single beams (indicating that the steeper slope is facing the beam), and $\bar{\epsilon}\approx 0.01$ for opposed beams. 
We also checked, that the skewness of the $\epsilon$-distribution is reduced by using opposed beams. Thus, this setup may have merits in producing ripple structures of higher quality and order.

\subsection{\label{sec:Crossed}Crossed ion beams}
Setups of multiple ion beams incident from different directions have been proposed by 
Carter in Refs~.\onlinecite{carter04} and \onlinecite{carter05}. Vogel and Linz developed a continuum theory for a setup with four beams , 
all with the same polar angles of incidence and crossing with $\Delta\phi=90^\circ$. Their theory
is based on a damped Kuramoto-Sivashinsky equation and takes into 
account terms up to fourth order in a gradient expansion of the erosion term. 
They predict square patterns, which can be stable in the long time regime under specific conditions.\cite{vogel07}
Joe et al. performed experiments with two ion-beams with $\theta=73^\circ$ and  $\Delta\phi=90^\circ$ on Au(001).
\cite{joe07,joe09} They find nanodot patterns with square symmetry (albeit with a rather low degree of order),
if the fluxes of the two beams are precisely balanced. Otherwise, the structures develop into modulated ripples. 

\begin{figure}
\begin{flushright}
(a)\includegraphics[width=.271\linewidth]{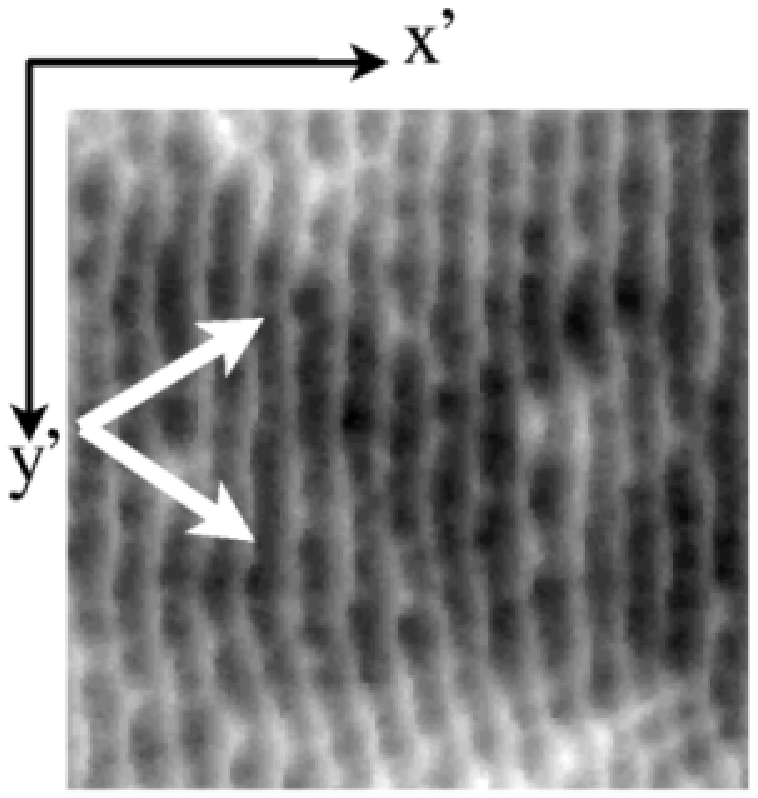} (b)\includegraphics[width=.25\linewidth]{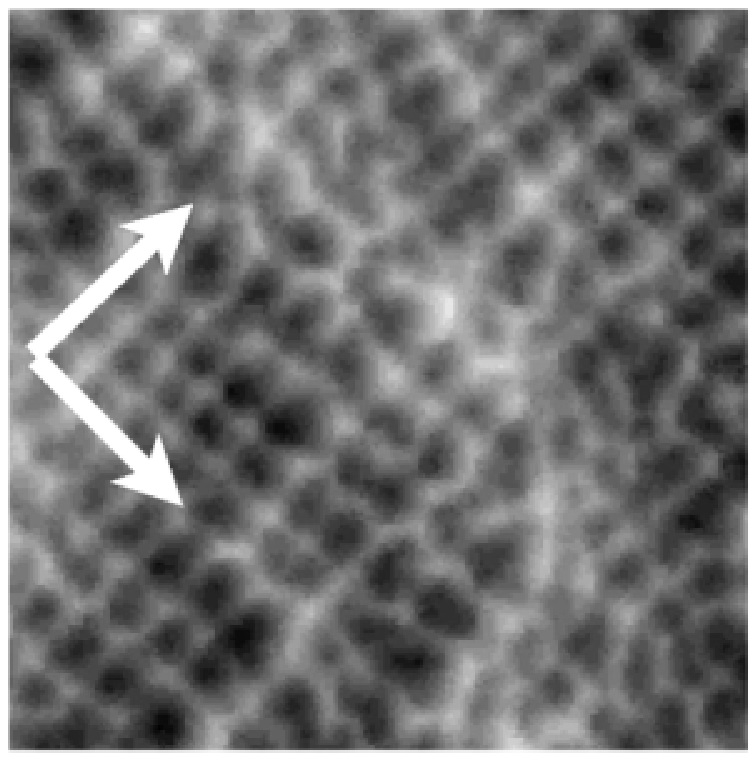} 
(c)\includegraphics[width=.25\linewidth]{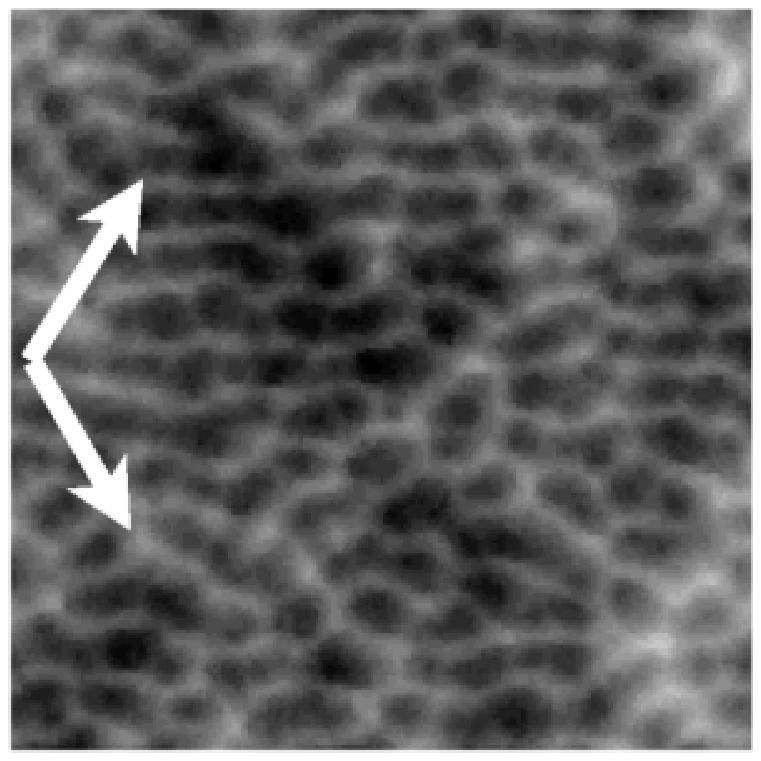}

(d)\includegraphics[width=.25\linewidth]{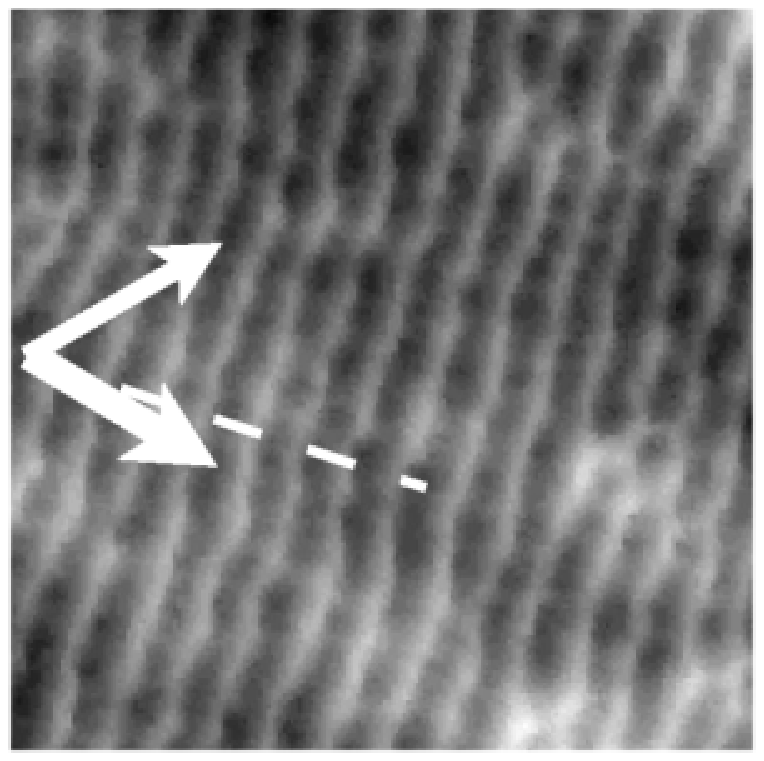} (e)\includegraphics[width=.25\linewidth]{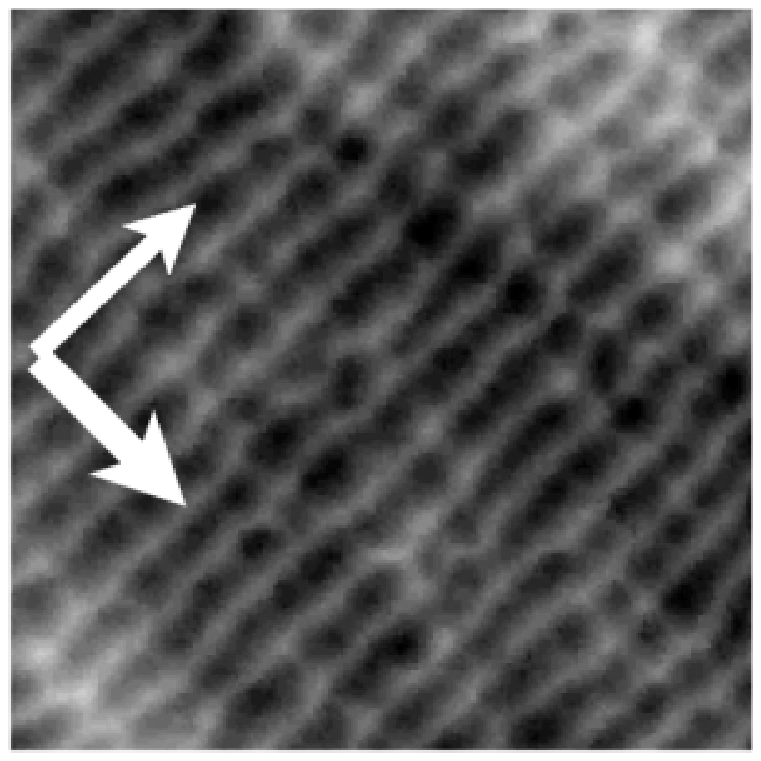} 
(f)\includegraphics[width=.25\linewidth]{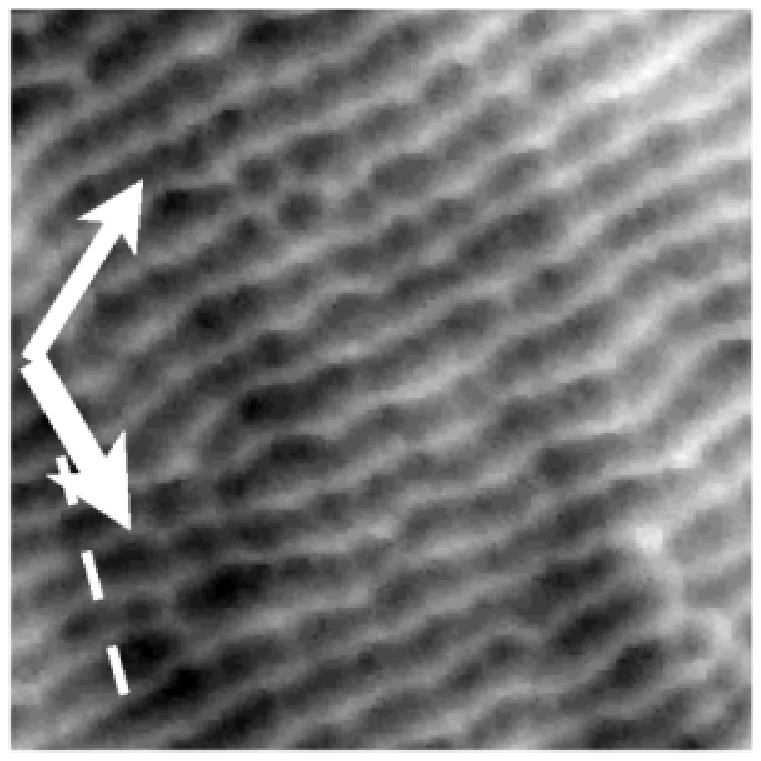}

(g)\includegraphics[width=.273\linewidth]{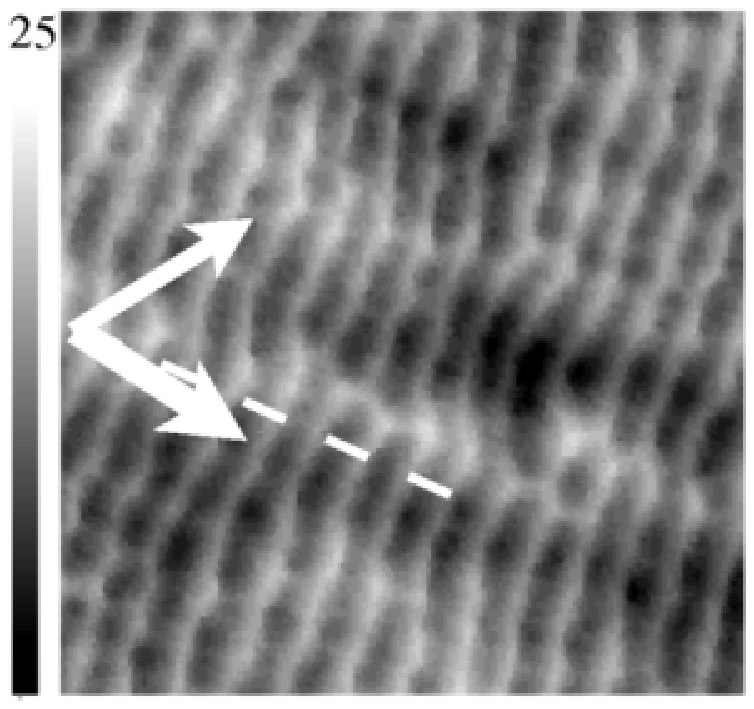} (h)\includegraphics[width=.25\linewidth]{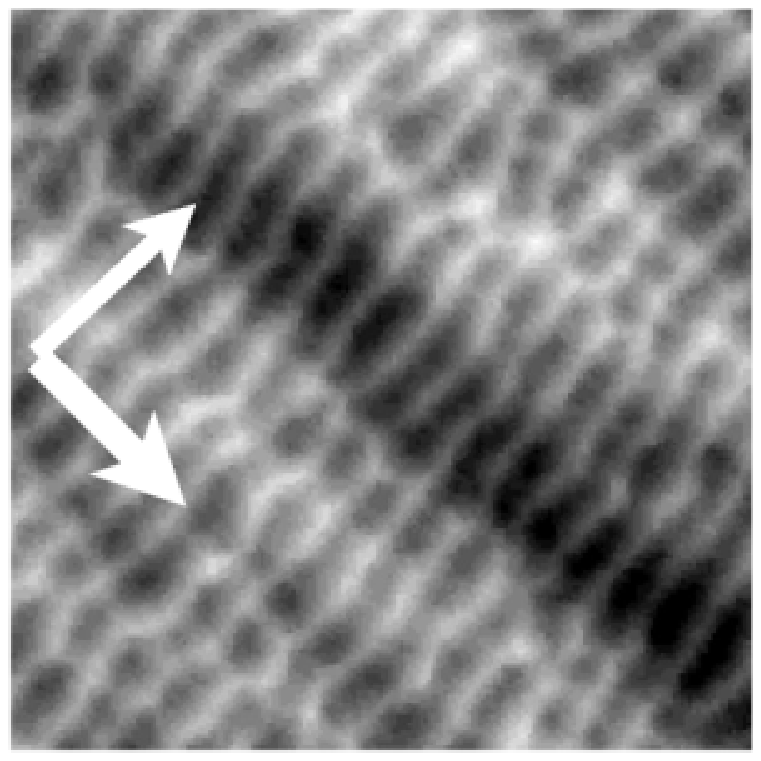} 
(i)\includegraphics[width=.25\linewidth]{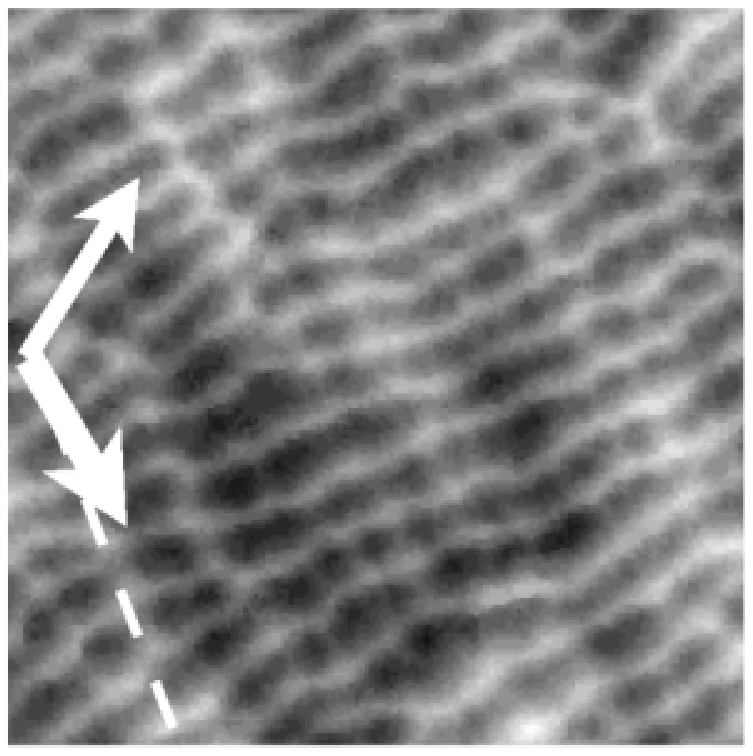}                                                    \end{flushright}
\caption{\label{fig:dual} \textit{upper row}: Rippled surfaces after $40 ML$ sputtering by
  two balanced ion beams incident from $\theta=50^\circ$ 
and separated in azimuthal angle by (a)$\Delta \phi =60^\circ$,  (b) $\Delta \phi=90^\circ$ and (c) $\Delta \phi=120^\circ$.
 \textit{middle row}:  same as (a)-(c) but for imbalanced ion beams,
 which differ in intensity by a factor of 2 ($f=1/2$) corresponding to beams (2) and (3) in Table~\ref{tab:table1}. \textit{lower row}: same as \textit{middle row}, here
 the imbalance is generated by different beam parameters, (1) and (4) in Table~\ref{tab:table1}.  
Arrows indicate the directions of ion-beams. In middle and lower row  bigger
arrows correspond to the dominant ion beam. 
Dashed lines indicate the directions predicted for the wavevectors of ripples by linear theory (see main text).}
\end{figure}

In Fig.~\ref{fig:dual} upper row, we show simulation results for perfectly balanced ion beams (beam parameters (1) in Table~\ref{tab:table1}). The middle row of Fig.~\ref{fig:dual} shows results for ion beams with different intensities incident from the same polar
angle ((2) and (3) in Table~\ref{tab:table1}) and the lower row shows corresponding results for ion beams of identical intensities, but
incident from different polar angles and with different beam
parameters ((1) and (4) in Table~\ref{tab:table1}).
White arrows in  Fig.~\ref{fig:dual} indicate the projection of the 
ion beams into the $x-y$ plane (thus they enclose $\Delta\phi$), and we have chosen the geometry such that the x-axis always is the bisector of this angle. 
For balanced beams and $0\leq \Delta\phi< 90^\circ$, we observe ripples with a wave vector parallel to the  x-axis, for $90^\circ<\Delta\phi<180^\circ$, the ripple wave vector
is parallel to the y-axis, but the ripple pattern shows more defects. Exactly at $\Delta\phi=90^\circ$, square patterns replace the ripples. 
For unbalanced beams, we observe ripples in oblique directions, which we will discuss below. The erosion rates parallel and
perpendicular to the beam projection onto the $x-y$ plane, $\nu_\parallel$ and $\nu_\perp$, are given in Table~\ref{tab:table1}. 
The middle and lower row of Fig.~\ref{fig:dual} are given to emphasize
that we did not observe any differences in pattern formation due to
different mechanisms of imbalance. All mechanisms we tried lead to the
same patterns, if they imply the same erosion rates of linear
theory, up to a common constant factor.      

Note, however, that in all our simulations both rates $\nu_\parallel$ and $\nu_\perp$ are
negative --- unlike the situation in
Ref.~\onlinecite{joe09} --- indicating that the crossed beams do not mutually suppress their generated ripples.  Thus we have nothing to say here about the puzzling results of Ref.~\onlinecite{joe09}, who find structures in DIBS setups, for which a straightforward application of continuum theory would predict smooth surfaces. 

Let us analyze our findings within the simple framework of linear Bradley-Harper theory. Suppose beams 1 and 2  are characterized by erosion rates
$\nu^{(b)}_{A}, \, b=1,2$ and $A=\parallel, \perp$.  Averages of the rates over the two beams are denoted by $\bar{\nu}_{A}=(\nu^{(1)}_{A}+\nu^{(2)}_{A})/2$ 
and we introduce $\Delta\nu_{A}=(\nu^{(1)}_{A}-\nu^{(2)}_{A})$. 
According to linear theory, the growth of Fourier modes $|h(k_x, k_y,t)|\propto\exp(\Gamma t)$ is controlled by the (real) growth rate $\Gamma(k_x,k_y)$, 
which is a quadratic form of the wave vector, i.e. $\Gamma=\mb{k}^t\hat{\nu}\mb{k}$ with a matrix $\hat{\nu}$ of erosion rates.  
The growth rate of the fastest growing mode and its direction are
obtained by determining the largest eigenvalue of $\hat{\nu}$ and the corresponding eigenvector. $\hat{\nu}$ is easily calculated. Its matrix elements take on the
form
 \begin{eqnarray}
 \label{eq:rates}
 \hat{\nu}_{xx} & =& 2(\bar{\nu}_{\parallel}+\bar{\nu}_{\perp})+(\bar{\nu}_{\parallel}-\bar{\nu}_{\perp})\cos\Delta\phi\\\nonumber
 \hat{\nu}_{yy} & =& 2(\bar{\nu}_{\parallel}+\bar{\nu}_{\perp})-(\bar{\nu}_{\parallel}-\bar{\nu}_{\perp})\cos\Delta\phi\\\nonumber
 \hat{\nu}_{xy} & = &  (\overline{\Delta\nu}_{\parallel}-\overline{\Delta\nu}_{\perp})\sin\Delta\phi,
 \end{eqnarray} 
 and $\hat{\nu}_{yx}=\hat{\nu}_{xy}$. For balanced beams, the $\Delta\nu_{A}$ vanish and the erosion rate matrix becomes diagonal, indicating that ripples 
 will only appear with wave vectors either parallel to the x-axis or parallel to the y-axis. At $\Delta\phi=90^\circ$, the rates in both directions become degenerate and square 
 patterns will emerge, if they are stabilized by the nonlinear terms. As $\cos(\Delta\phi+\pi/2)=-\cos\Delta\phi$, it is obvious from Eq. \ref{eq:rates} that the regime
  $90^\circ<\Delta\phi<180^\circ$ can be mapped to $0<\Delta\phi<90^\circ$ by interchanging x and y. This explains the main features of the upper row of  Fig.~\ref{fig:dual}.

\begin{figure}
\includegraphics[width=\linewidth]{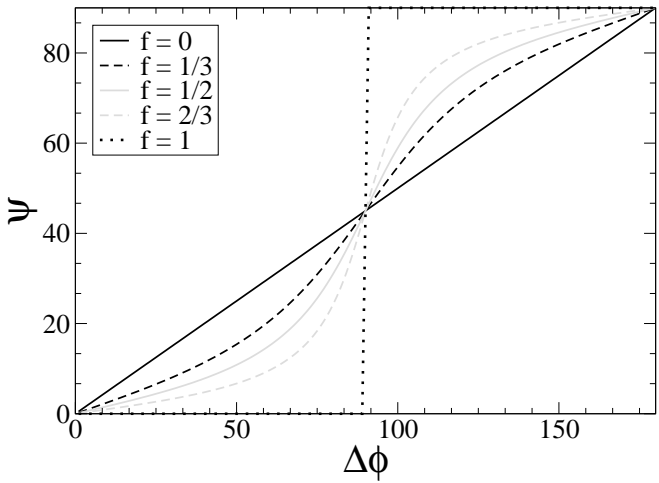}
\caption{\label{fig:say-df} Orientation of fastest growing mode for
  different values of $\Delta \phi$. $\psi$ denotes the angle of the
  ripple wave vector with the x-axis. Different lines correspond to different imbalances f (see main text).}
\end{figure}

For unbalanced beams, the orientation of ripples will generally depend both on $\Delta\phi$ and the imbalances in growth rates $\Delta\nu_{\parallel}, \Delta\nu_{\perp}$, 
but for $\Delta\phi=90^\circ$ the situation is simpler. The cosine terms vanish, and the eigenvectors $\mb{e}_{\pm}\propto (1, \pm 1)$ 
of the simpler matrix become independent of the rate imbalances. Thus,
the linear theory predicts ripples with wavevectors inclined by $\psi=45^\circ$ or
$\psi=135^\circ$ with respect to the x-axis, i.e. parallel to one of
the beams (from the eigenvalues it follows that the wave vector is
parallel to the dominant beam), irrespective of the amount or nature
of the imbalance in growth rates, if $\Delta\phi=90^\circ$. This
result is in accordance with the experimental finding in Ref.~\onlinecite{joe09}.
The directions $\psi$ for other values of $\Delta\phi$ are easily obtained, if the 
imbalance is known. In Fig.~\ref{fig:dual}, we have indicated these directions with dashed white lines and find an overall satisfactory agreement 
of simulation results with this prediction. Fig.~\ref{fig:say-df} shows the predicted deviations of the orientation of ripple wavevectors from the x-axis, i.e. the bisector of the beam directions for different ratios of rates, $f=\nu^{(1)}_{A}/\nu^{(2)}_{A}$.

\begin{table}
\caption{\label{tab:table1}Beam parameters for our DIBS setup}
\begin{ruledtabular}
\begin{tabular}{ccccc}
& $\theta$ & $a$ & $\nu_\parallel$ & $\nu_\perp$\\
\hline
(1)& $50^\circ$&9.3&-4.4&-0.86\\
(2)& $50^\circ$&9.3&-5.8&-1.14\\
(3)& $50^\circ$&9.3&-2.9&-0.57\\
(4)& $35^\circ$&7.3&-1.5&-0.81\\
\end{tabular}
\end{ruledtabular}
\end{table}

\subsection{\label{sec:SIBS}Sputtering of rippled surfaces}
 Since the possibilities of simultaneous sputtering by multiple beams are often limited 
in experiments, sequential sputtering by a single beam from different directions seems to be a more promising setup. 
Vogel and Linz proposed SIBS as a general substitute for multi-beam sputtering.\cite{vogel07} Note that in  SIBS setups,  
a precise balance between fluxes of multiple beams, which may be difficult to achieve in DIBS, can be adjusted by tuning the exposure time in each direction.  

Joe et al. \cite{joe09} have recently performed experiments on Au(001) using an ion beam     
(incident from $\theta=72^\circ$),  for which ripples with wave vectors perpendicular to the projection of the beam direction 
into the initial surface plane are formed. 
After rotating the target by $90^\circ$  (keeping $\theta$ fixed), the initially formed ripples are very rapidly 
destroyed and new ripples build up in the correspondingly rotated direction, but the authors could not get patterns corresponding to a superposition 
of two generations of ripples as expected from the predictions of continuum theory.  

\begin{figure}
(a)\includegraphics[width=.43\linewidth]{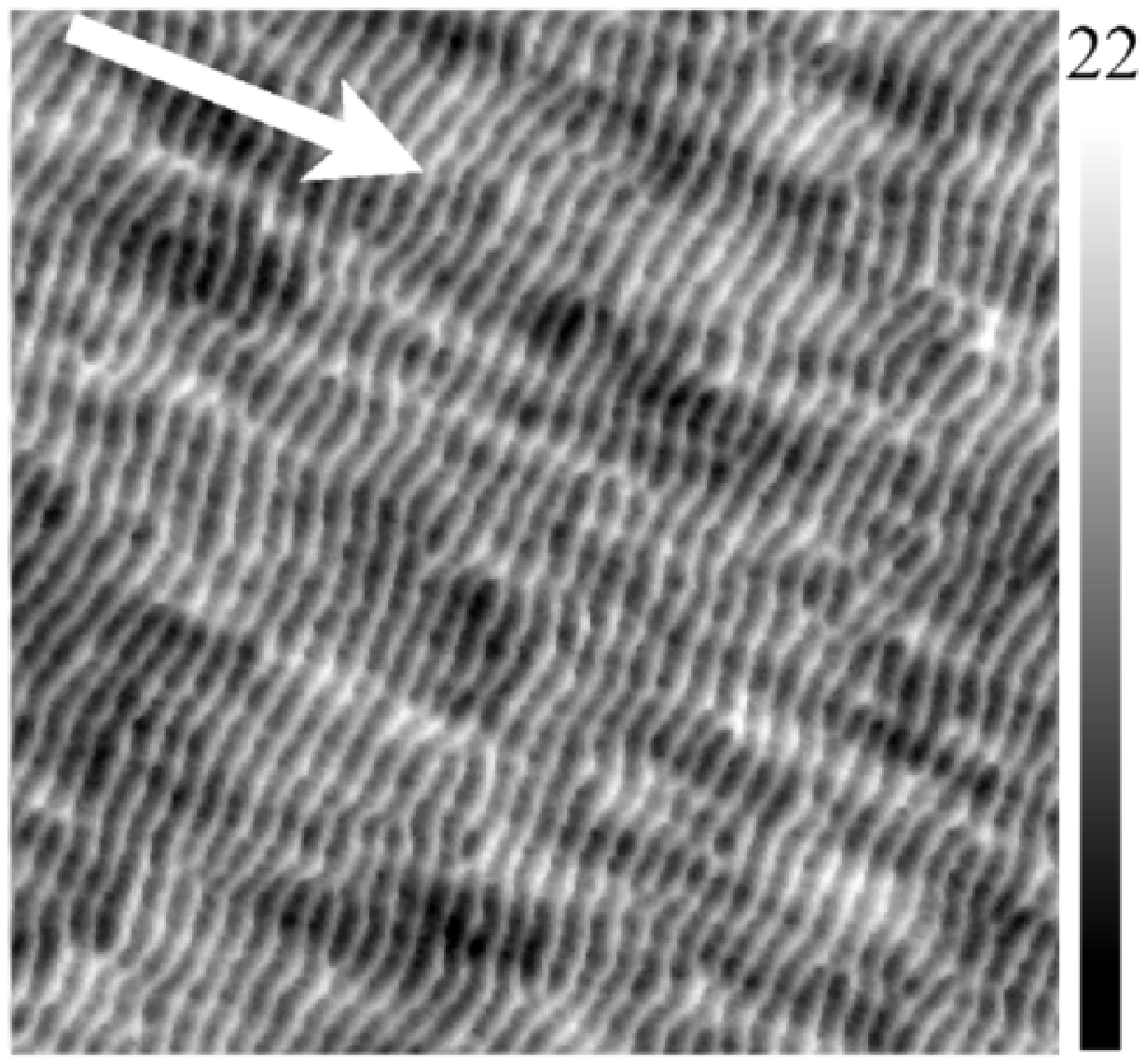} (b)\includegraphics[width=.43\linewidth]{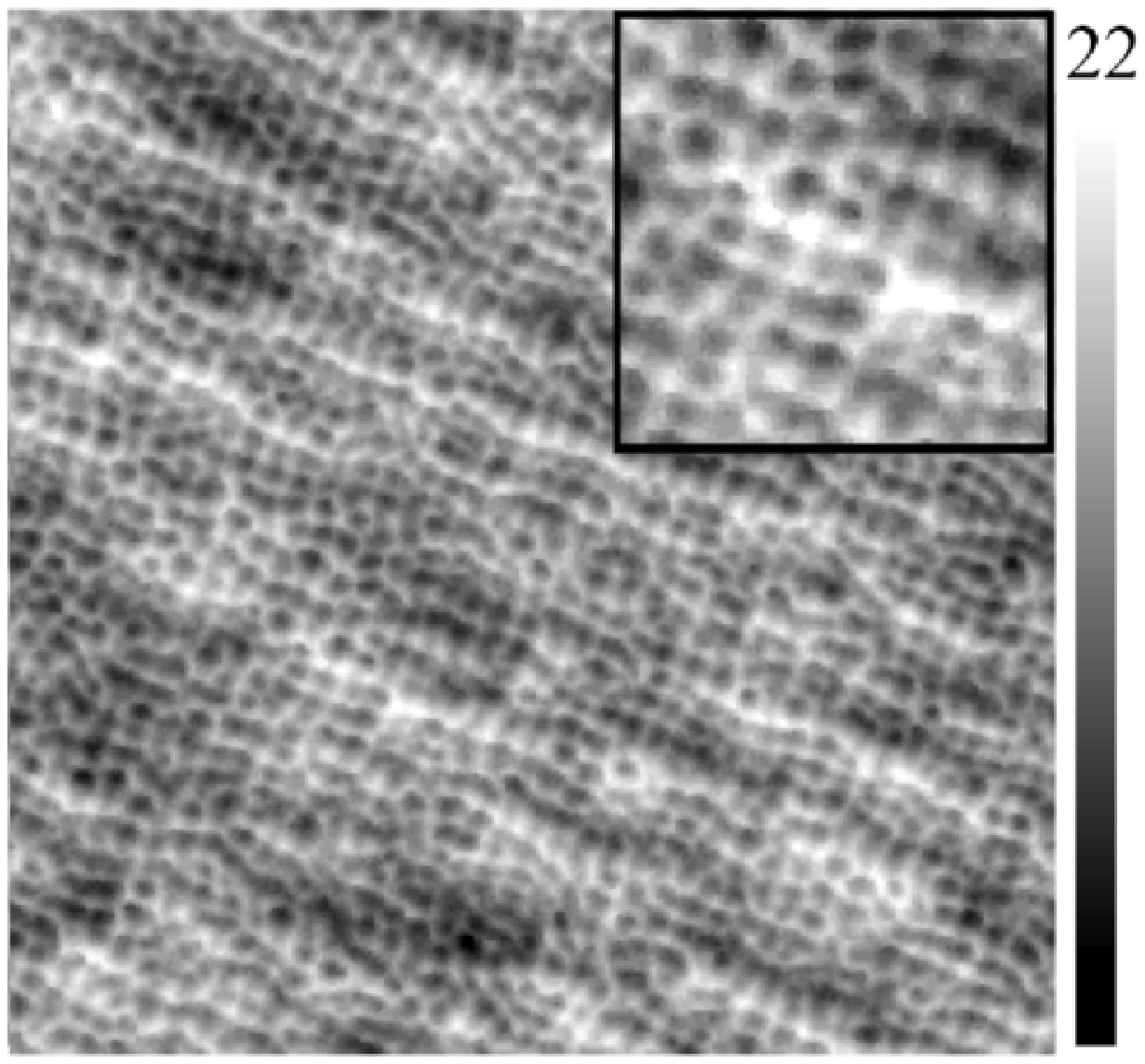}

(c)\includegraphics[width=.43\linewidth]{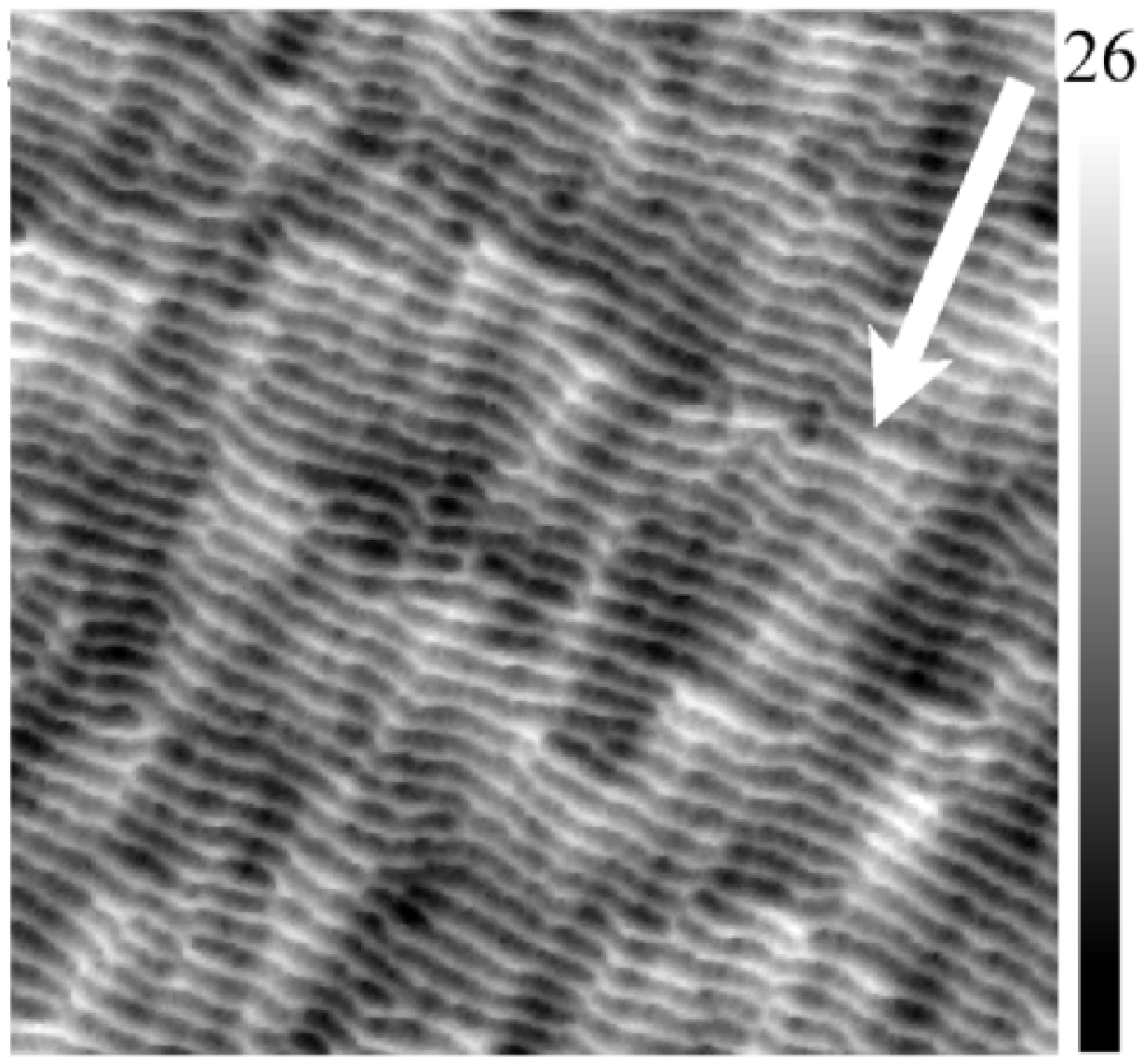} (d)\includegraphics[width=.43\linewidth]{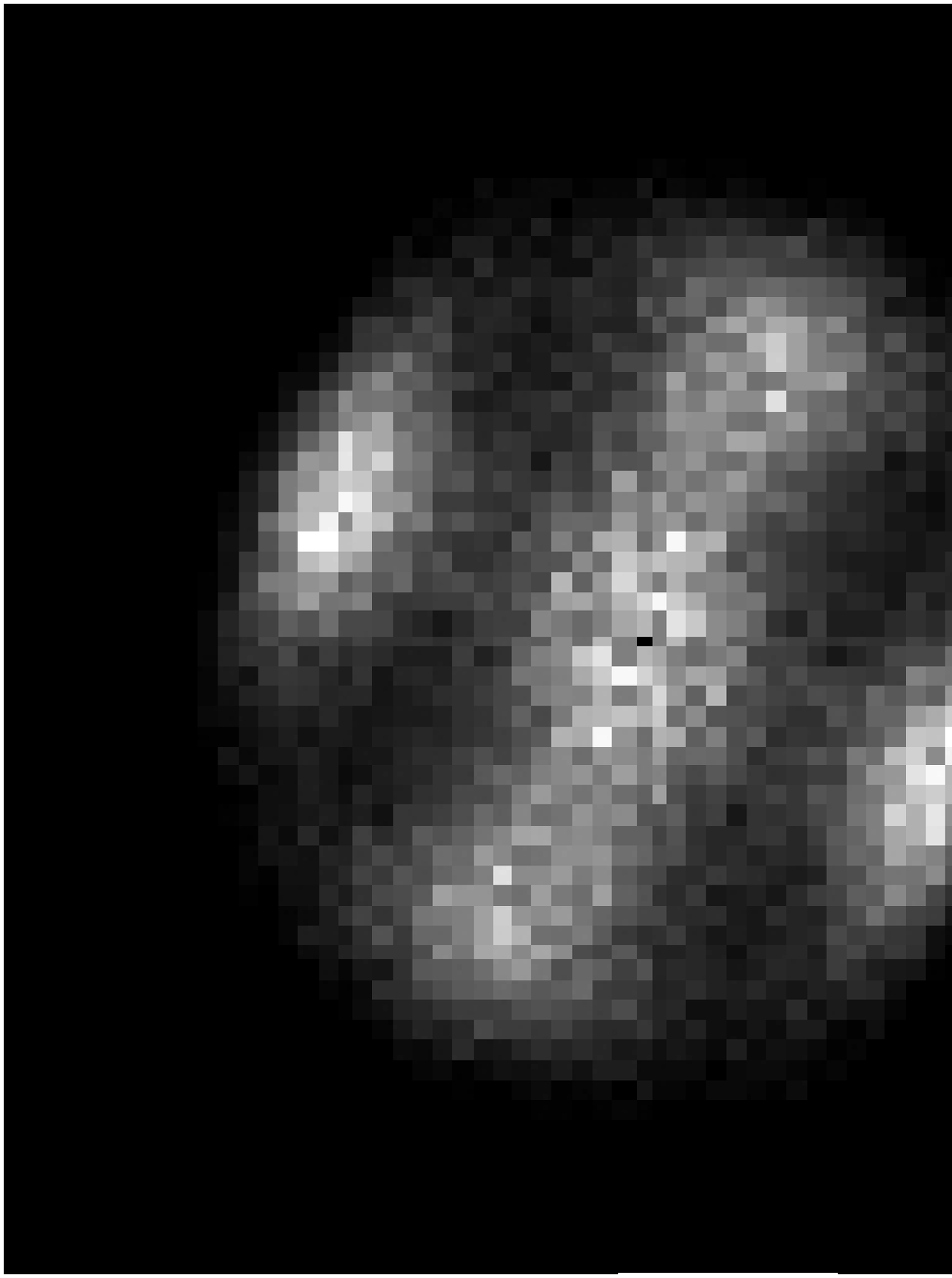}
\caption{\label{fig:sibs} Snapshots of surface profile during SIBS.  At $t=9$, a
$\Delta\phi=90^\circ$ rotation step occurs. 
Shots are at (a) $t=9$, (b) $t=10.8$ (inset: zoomed in) and (c) $t=18$. Arrows indicate
the direction of ion beams. 
In panel (d) the structure factor $|h(k_x,k_y)|^2$ of the profile of panel (b) is shown.}
\end{figure}

We have simulated SIBS with ion beams incident from $\theta=50^\circ$ and a rotation step of $\Delta \phi= 90^\circ$ after $9$ monolayers of erosion.  
Initially, ripples appear with wave vectors parallel to the direction of the projected ion beam into the x-y plane (see  Fig.~\ref{fig:sibs}(a)). 
After the rotation step, a correspondingly rotated ripple pattern builds up as shown in Fig.~\ref{fig:sibs}(c) . In a narrow time window, 
shortly after the rotation step (shown in Fig.~\ref{fig:sibs}(b), at $t=10.8 ML$) a 
superposition of ripples of both orientations is observed. 
Fig.~\ref{fig:sibs}(d) shows the structure factor of the height profile, which depicts the degree of order. 

\begin{figure}
\includegraphics[width=\linewidth]{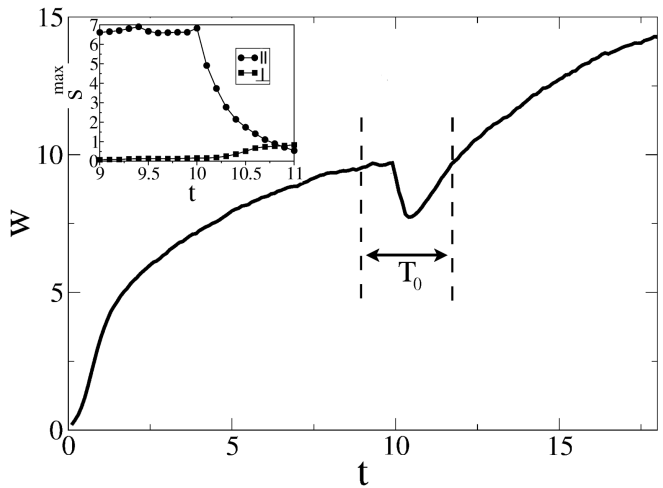}% Here is how to import EPS art
\caption{\label{fig:sibs-roughness}Roughness of surface patterned by SIBS vs. time. 
The surface is rotated by $90^\circ$ at $t=9$. The inset shows the
temporal evolution of the maximum of the structure factor $S_{max}=\max(|h(\vec{k})|^2)$ for wavevectors in $\parallel$ and in
$\perp$ direction.}
\end{figure}

Beyond $\simeq 1 ML$ of erosion after the rotation step the time evolution of surface roughness (shown in Fig.~\ref{fig:sibs-roughness}) differs strongly from what one expects
in linear continuum theory.  There, the height of prestructured
ripples, having already reached saturation, would neither grow
further, nor will it decrease. Ripples generated by the new beam
direction will grow exponentially and thereby catch up the height of
the prestructure. The different scenario we observe in the simulations
is shown in the inset of Fig.~\ref{fig:sibs-roughness}. After $\simeq 1 ML$,
during which the system follows the scenario predicted by
Bradley-Harper theory, the prestructure collapses very rapidly and is
flattened, leading to a transiently decreasing roughness. 
We define a structural relaxation time $T_0$ as the length of the
interval from the rotation step through the transient decrease of
roughness to the point, where roughness has retained its value
immediately before the rotation step. Roughly this time interval
contains all processes necessary to rotate the ripple pattern from the
previous to the current rotation step. In our case $T_0\simeq 2.5$.
Although $T_0$ cannot be obtained from linear rate theory our
observations need not be in contradiction with continuum theories, as
the prestructure constitutes an initial condition beyond the range of validity of linear theory.   
Let us reemphasize that the roughness decreases transiently, despite
the fact that both linear erosion rates $\nu_{\parallel}$ and $\nu_{\perp}$ (see Table (\ref{tab:table1})) indicate
an unstable growth of fluctuations in x and y direction.
A superposed square pattern is observable, if the height of the growing new ripples and the shrinking old ones
become comparable. For our parameters this takes place at $\sim 1.5 ML$ after rotation in a narrow time window.  
In the experiments by Joe et al.\cite{joe09} values of growth rates are 
$\nu_x=14.2$ and $\nu_y=-0.9$. Since $\nu_x>0  $ and $\nu_y<0$, linear rate theory predicts that the old generation of ripples will decay immediately after a rotation step, which may cause an even faster collapse of the prestructure and make the
time window of existence of a superposed structure unobservably small.

\subsection{\label{sec:Rotation}Sputtering of continuously rotating sample}
Sample rotation during IBS is applied for various reasons.  
One motivation is to achieve suppression of pattern formation in SIMS and AES depth profiling.\cite{zalar85,zalar86}
There are numerous reports that RIBS can suppress surface roughening and enhance the resolution of depth profiling (see Ref.~\onlinecite{carter98} and its references).
Although this method is frequently used,  there are very few systematic studies of the effects of different parameters, in particular of the angular velocity of rotation.\cite{tanemura92} Previous discrete simulations could not successfully 
 explain the observed strong suppression of roughening. \cite{koponen97}
Recently, IBS has become a popular method for smooth etching 
of metallic surfaces.\cite{reichel07} Here too, sample rotation has been proposed as a 
practical measure \cite{reichel07}  to prevent nano-scale roughening.
A different motivation to use RIBS is the control of pattern formation.
Frost et al. have found that off-normal IBS 
with sample rotation may lead to formation of hexagonal, close packed quantum dots.\cite{frost00,frost03} 
This is attributed to a restoration of axial rotation symmetry with respect to the average surface normal, which is broken by off-normal incidence of a single ion beam at fixed azimuthal angle.  
Dot formation in rotated, off-normal IBS has been found in the framework of continuum theories 
\cite{bradley96a,frost02,castro05},  assuming a flux of 
incoming ions, which is distributed evenly over all azimuthal angles. This assumption corresponds to the limit of high rotation frequencies. Dots also appear in MC simulations performed in the high rotation frequency limit for a wide range of parameters.\cite{yewande07}

In the present work, we focus on the systematic dependence of height fluctuations on the rotation frequency, irrespective of the random or deterministic nature of these fluctuations. Therefore, we will study
the roughness $w=\langle(h-\langle h\rangle)^2\rangle$, averaged over the sample and an ensemble, as function of time and rotation frequency $\omega$. 
Reported rotation frequencies cover a range from 0.1 to 15 rpm for different fluxes and different types of ions and materials. 
\cite{frost03,zalar85,zalar86,konarski95,cui05}
There is a predictions of the scaling of height with $\omega$, which, in Ref.~\onlinecite{bradley96b},  
is given in the form
 \begin{equation}
 \label{eq:scale} 
h(\vec{k},t) \propto \exp(\frac{(\nu_\perp-\nu_\parallel)}{4\omega}k^2\sin(2\omega t)).
 \end{equation} 
It is based on the original linear Bradley-Harper theory in rotating coordinate systems. 
Furthermore, Cui et al. reported that ripples do not form  for
angular frequencies greater than 0.1 rpm.\cite{cui05} (for a flux of $3.5 \times 10^ {14}$ ions cm$^{-2}$ s$^{-1}$ of 300 eV Ar$^+$ on GaN substrate). 

\begin{figure}
\includegraphics[width=\linewidth]{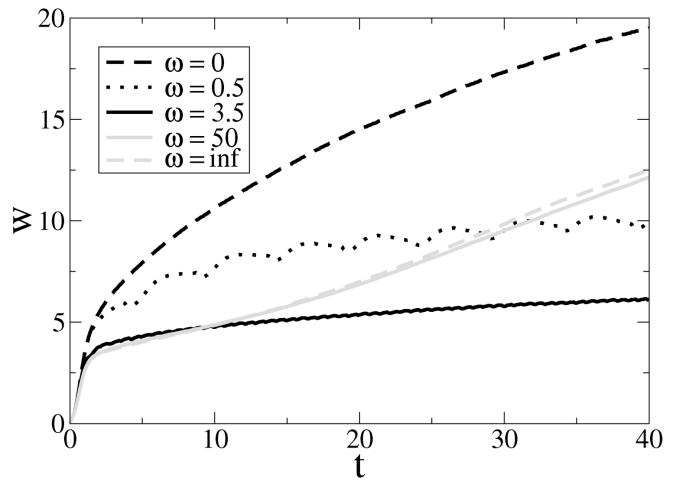}% Here is how to import EPS art
\caption{\label{fig:w-rough}Time evolution of surface roughness during
  RIBS  for different rotation frequencies.}
\end{figure}

\begin{figure}
\includegraphics[width=\linewidth]{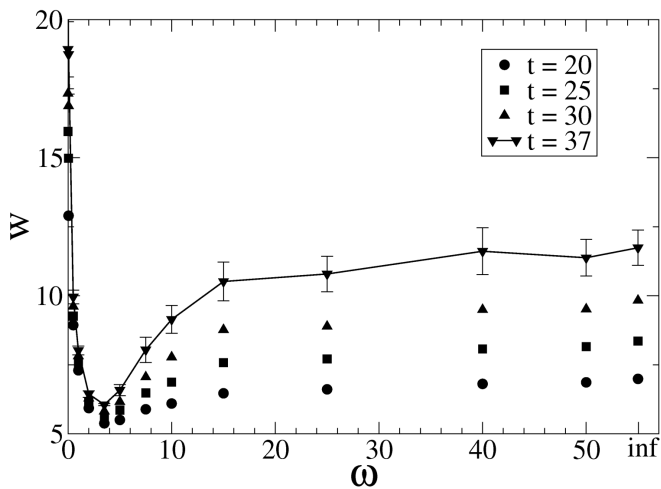}% Here is how to import EPS art
\caption{\label{fig:w-omega}Roughness against angular frequency
  $\omega$ at different times.}
\end{figure}

In section~\ref{sec:SIBS}, we have defined a characteristic response time $T_0(\Delta\phi)$ of the IBS generated structures to sudden changes $\Delta\phi$ of the beam direction. 
We propose that this time scale is also of relevance for RIBS, as surface structures might follow rotation frequencies much smaller than $\omega_0=\Delta \phi / T_0$ adiabatically. On the other hand $\omega\gg \omega_0$ might correspond to the high frequency limit.  
We performed simulations with different rotation frequencies varying in the range $0.05 \cdots 50\, \omega_0$, with $\omega_0\simeq 36^\circ$ \textit{per eroded monolayer} 
taken from our SIBS simulations.
For low frequencies, ripples form and rotate in synchrony with the beam direction. For high frequencies,
($\omega = \infty$ is included as it corresponds to random azimuthal directions of incoming ions chosen from a flat distribution), 
cellular structures of growing size are observed.
These findings are in agreement with predictions in Ref.~\onlinecite{bradley96a}. 
A mixture of short ripples and cellular structures appears at intermediate $\omega$. 
Somewhat surprisingly, the roughness of the surface is not a monotonous function of rotation frequency. 
In Fig.~\ref{fig:w-rough} roughness  is shown as a function of time for different  $\omega$.
 For low (including $\omega = 0$) and high frequencies, it grows monotonically, approaching approximately equal growth rates beyond $t\approx 30$, 
independent of $\omega$. For intermediate $\omega$, the roughness shows oscillations (with frequencies $\Omega\approx 4\omega$), and its growth rate is strongly suppressed. 
For $\omega=3.5\,\omega_0$ growth rate reaches a minimum. 
In Fig.~\ref{fig:w-omega} the roughness, averaged over a period $T=2\pi/\omega$, is shown against $\omega$,  for times up to $t=37$ . For increasing $t$, the minimum at $\omega\simeq 3.5\, \omega_0$ becomes more and more pronounced.
Thus our simulations predict an optimal 
rotation frequency, if preparations aim at smooth surfaces. This frequency also marks the frequency scale beyond which ripples do no longer appear. The very rapid crossover to non-ripple structures is in accordance with findings of Ref.~\onlinecite{cui05}.

\begin{figure}
\includegraphics[width=\linewidth]{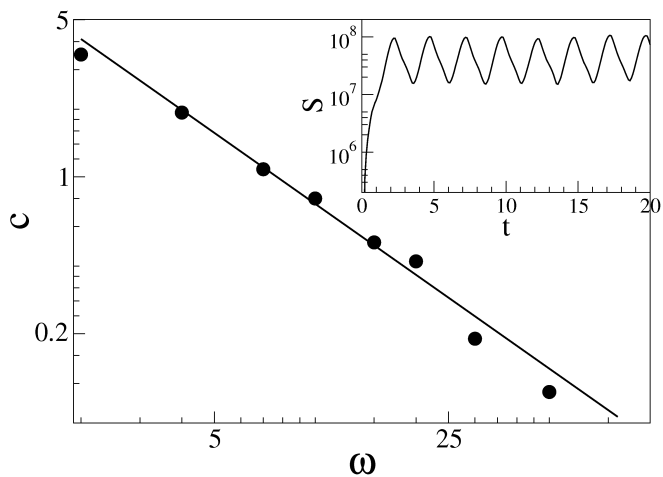}% Here is how to import EPS art
\caption{\label{fig:scale} The logarithm of the oscillation amplitude of structure factor for a given $k$, vs. $log\omega$. \textit{Inset:} Time evolution of structure factor for a given $k$, for $\omega=2\omega_0$ displays the oscillations. The amplitude of these oscillations is shown in the main figure.}
\end{figure}

To compare our results with the prediction of Eq.~\ref{eq:scale} about the scaling of height with $\omega$, we studied
$ S=|h(\vec{k},t)|^2 $ for a fixed value of wavevector $\vec{k}$.
 It grows rapidly and then oscillates with frequency
$2\omega$ around a saturation value with an $\omega$-dependent amplitude $c$, as shown in the inset of Fig.~\ref{fig:scale}.
The oscillatory behavior with frequency $2\omega$ is also present in Eq.~\ref{eq:scale}. The main part of the figure depicts the decrease of the oscillation amplitude with increasing $\omega$ in a double logarithmic plot. The fitted line has a slope of $-1.05\pm0.05$, which is very close to the $1/\omega$ behavior suggested by Eq.~\ref{eq:scale}.  Note, however, that this equation was derived from linear theory and has to be multiplied by an exponential growth factor, whereas our result applies to the non-linear saturation regime. Due to the rapid initial growth, the regime of validity of Eq.~\ref{eq:scale} remained unobservable.

\begin{figure}
\includegraphics[width=\linewidth]{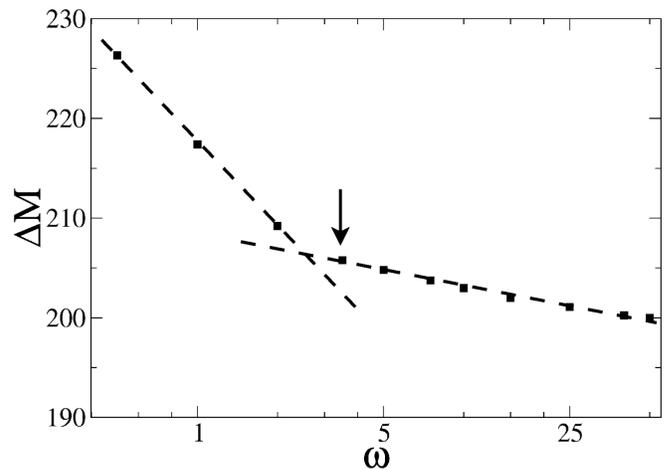}% Here is how to import EPS art
\caption{\label{fig:yield} Total amount of eroded material (integrated yield) after $40 ML$ of sputtering of rotating samples vs. rotation frequency. The arrow is at $\simeq3.5 \omega_0$, the frequency which minimize the total roughness.}
\end{figure}

We also measured the total amount of eroded material, $\Delta M$ up to $t=40$ for different rotation frequencies.  This integrated yield decreases with increasing $\omega$ and displays a clear distinction between a low- and a high-frequency regime in a semi-log plot, shown in Fig.~\ref{fig:yield}. The interpolated crossover frequency between these regimes is very close to $\simeq 3.5\omega_0$, the frequency, which minimizes the total roughness (see Fig.~\ref{fig:w-omega}). This is in accordance with our findings (see Fig.~\ref{fig:sibs-roughness}) that more pronounced rippled prestructures (developing at low rotation frequencies at an angle $\omega t$) decay more rapidly (at $t+dt$) than the smaller and less regular height fluctuations, which appear at higher rotation frequencies.

\section{\label{sec:Conclusion}Conclusion and Outlook}
We studied dual ion beam sputtering (DIBS), sequential ion beam sputtering (SIBS) and 
rotating ion beam sputtering (RIBS) by a kinetic MC simulation
technique, which combines erosion events due to single ions and
surface diffusion. For a DIBS setup with two diametrically opposed
beams, we did not confirm predictions by Carter \cite{carter05}, but
rather found non-moving ripples with orientations as in a single ion
beam setup. The ripples
have a higher degree of order and more symmetrical slopes as compared
to those created by single beam sputtering. For DIBS setups with
crossed ion-beams, we find ripple patterns for crossing angles
$\Delta\phi\neq 90^ \circ$ and square patterns for crossing of
balanced beams at exactly right angle. The ripple orientations follow
the predictions from linear Bradley Harper theory. Any kind of
beam-imbalance leads to ripple patterns oriented according to the
dominant beam. This is in accordance with the experimental observation
in Ref.~\onlinecite{joe09}. For SIBS setups, we found a vary rapid
destruction of the ripple prestructure of the 
previous rotation step, which cannot be explained by linear
Bradley-Harper theory.  The flattening of the prestructure leads to a
transient decrease in total roughness.  Only within a very short time
window, the growing new generation of ripples and the shrinking old
ones lead to a superposed square pattern. Thus we could not confirm
propositions to use SIBS as a universal substitute for complicated
multi-beam setups. The rapid destruction of the prestructure is in
accordance with findings of Ref.~\onlinecite{joe09}. For RIBS setups we
observed a non-monotonic dependence of roughness upon rotation
frequency. At a frequency scale set by the structural relaxation time of prestructures,
which can be observed in SIBS simulations, an increasingly pronounced
minimum of roughness occurs with increasing time. We found that the structure factor at fixed wavevector  rapidly approaches stationary oscillations around a saturation value with oscillation amplitudes inversely proportional to frequency. This behavior was also predicted from linear theory, but seems to have a much broader range of validity.

\begin{acknowledgments}
We like to thank Roland Bennewitz, Hans Hofs\"ass and Kun Zhang for useful discussions.
\end{acknowledgments}

\newpage %Just because of unusual number of tables stacked at end
\bibliography{rotation}% Produces the bibliography via BibTeX.

\end{document}